\documentclass[reprint, amsmath,amssymb, aps, showpacs]{revtex4-1}
\usepackage{graphicx}
\usepackage{dcolumn}
\usepackage{bm}
\usepackage{amssymb}   
\usepackage{color, colortbl}
\usepackage{textcomp}
\usepackage{ulem}
\usepackage{floatrow}
\usepackage{multirow}
\usepackage{hyperref}
\usepackage{natbib}
\usepackage{verbatim} 

\begin{document}

\title{Vacancies and oxidation of 2D group-IV monochalcogenides}

 \author{L\'idia C. Gomes}
 \affiliation{ Centre for Advanced 2D Materials and Graphene Research Centre, National University of Singapore, 6 Science Drive 2, 117546, Singapore }%
 \author{A. Carvalho}
 \affiliation{ Centre for Advanced 2D Materials and Graphene Research Centre, National University of Singapore, 6 Science Drive 2, 117546, Singapore }%
 \author{A. H. Castro Neto}
 \affiliation{ Centre for Advanced 2D Materials and Graphene Research Centre, National University of Singapore, 6 Science Drive 2, 117546, Singapore }%
 \date{\today}

\begin{abstract}
Point defects in the binary group-IV monochalcogenide monolayers of SnS, SnSe, GeS, GeSe
are investigated using density-functional-theory calculations.
Several stable configurations are found for oxygen defects, however we give evidence that
these materials are less prone to oxidation than phosphorene, with which monochalcogenides are isoelectronic and
share the same orthorhombic structure. 
Concurrent oxygen defects are expected to be vacancies and substitutional oxygen. 
We show that it is energetically favorable oxygen be incorporated into the layers substituting for a chalcogen ($\rm O_{S/Se}$ defects), and different from most of the other defects investigated, this defect preserves the electronic structure of the material.
Thus, we suggest that annealing treatments can be useful for the treatment of functional materials where loss mechanisms due to the presence of defects are undesirable.
\end{abstract}

\pacs{Valid PACS appear here}
\maketitle


\hyphenation{ALPGEN}
\hyphenation{EVTGEN}
\hyphenation{PYTHIA}

\section{Introduction}

Layered group-IV monochalcalgenides has become an important group of materials within the ever-growing family of two-dimensional crystals. Among the binary IV-VI compounds, SnS, SnSe, GeS and GeSe form a sub-group with orthorhombic structure belonging to the space group $D^{16}_{2h}$. Even though bulk structural, electronic and optical properties of these materials have been investigated since the 70's~\cite{PhysRevB.16.1616, nc.39.709, krist.148.295, PhysRevB.41.5227}, more recently their photovoltaic properties have been gaining considerable attention due to the increasing demand for efficient energy conversion technologies~\cite{am402550s, ashley.arxiv.1507}. The optimal band gap for photovoltaic solar cells of the naturally occurring bulk SnS, also known as herzenbergite~\cite{PhysRevB.70.235114, PhysRevB.92.085406}, boosted experimental and theoretical research on this material in the last years.  

Additional interest in group-IV monochalcogenides arose with the advances in experimental techniques of production and manipulation of low-dimensional materials, paving an avenue for research in the 2D field.
While the most studied 2D materials are hexagonal, as graphene, phosphorene layers are orthorhombic,\cite{PhysRevLett.114.046801} 
and therefore became a paradigm of anisotropy in 2D.
Anisotropy has important consequences,
for example the enhanced thermoelectric effect arising from the fact that the preferential axes for heat and electronic
conduction are orthogonal.\cite{fei-NL-14-288}
Monolayer group-IV monochalcogenides are isoelectronic with phosphorene and share the same structure,
but a lower symmetry, and therefore are expected to show large spin-orbit splitting~\cite{PhysRevB.92.085406}
 piezoelectricity and high ionic dielectric screening~\cite{PhysRevB.92.214103}, all of them absent in phosphorene.
Experimental progress in growth 
and exfoliation has already resulted in the isolation of bilayers,
and the isolation of monolayer is expected.\cite{nn303745e, am402550s, jacs.5b08236}

In this article, we reveal yet another aspect in which group-IV monochalcogenides are
more promising than phosphorene: their resistance to oxidation.
In fact one of the hindrances to the research and use of phosphorene is its tendency to
oxidize.\cite{PhysRevLett.114.046801, PhysRevB.91.085407}
Exposed to air, few-layer samples degrade in less than one hour,\cite{doganov-arXiv:1412.1274}
eventually producing phosphorus oxide and phosphoric acid. 
Thus, phosphorene devices require immediate encapsulation in order to maintain their I-V characteristics.\cite{avsar-arXiv:1412.1191,zhu-NL-15-1883}
However, group-IV chalcogenides have stronger bonds and therefore are expected to be less prone to 
oxidation.
In this article, we use first principles calculations to investigate point defects in group-IV monochalcogenide monolayers. Section~\ref{O-defects} is dedicated to the study of chemisorbed oxygen defects. Subsequently, in Section~\ref{V-O-defects} we study the effects of intrinsic vacancy defects and substitutional oxygen. Conclusions are presented in Section~\ref{conclusions}.

\section{Methods}

We use first-principles calculations based on density-functional theory to obtain the electronic and structural properties of oxidized monolayer monochalcogenides. 
We employ a first-principles approach based on Kohn-Sham density functional theory (KS-DFT)~\cite{PhysRev.140.A1133}, as implemented in the {\sc Quantum ESPRESSO} code.~\cite{Giannozzi2009}. The exchange correlation energy was described by the generalized gradient approximation (GGA) using the PBE~\cite{PhysRevLett.77.3865} functional. Interactions between valence electrons and ionic cores are described by Troullier-Martins pseudopotentials~\cite{PhysRevB.43.1993}. 
The Kohn-Sham orbitals were expanded in a plane-wave basis with a cutoff energy of 70~Ry, and for the charge density, a cutoff of 280~Ry was used. The Brillouin-zone (BZ) was sampled using a $\Gamma$-centered 10$\times$10$\times$1 grid following the scheme proposed by Monkhorst and Pack~\cite{PhysRevB.13.5188}. 

We used periodic boundary conditions along the three dimensions. The layers are placed in the $x$-$y$ plane, with the $y$ axis parallel to the puckering direction, where atoms are arranged in a zigzag shape. Along the perpendicular $x$ axis, the atoms form an armchair configuration. In direction perpendicular to the layers, we used vacuum regions of 10~\AA{} between adjacent images. Convergence tests with greater vacuum thickness were performed, and the values used are enough to avoid spurious interaction between neighbouring images.

The isolated defects were modeled using  3$\times$3  supercells ($M_{18}C_{18}$, with $M$=Sn,Ge and $C$=S,Se). The unitcells with an adsorbed oxygen atom have, therefore, $M_{18}C_{18}$O chemical composition, while for vacancies and O substitutional the concentrations are $M_{18(17)}C_{17(18)}$ and $M_{18(17)}C_{17(18)}$O, respectively.

\section{Results}
\subsection{Oxygen defects}
\label{O-defects}

\subsubsection{Crystal Structure and Energetics}

\begin{figure*}[!htb]
    \centerline{
  \includegraphics[scale=1.05]{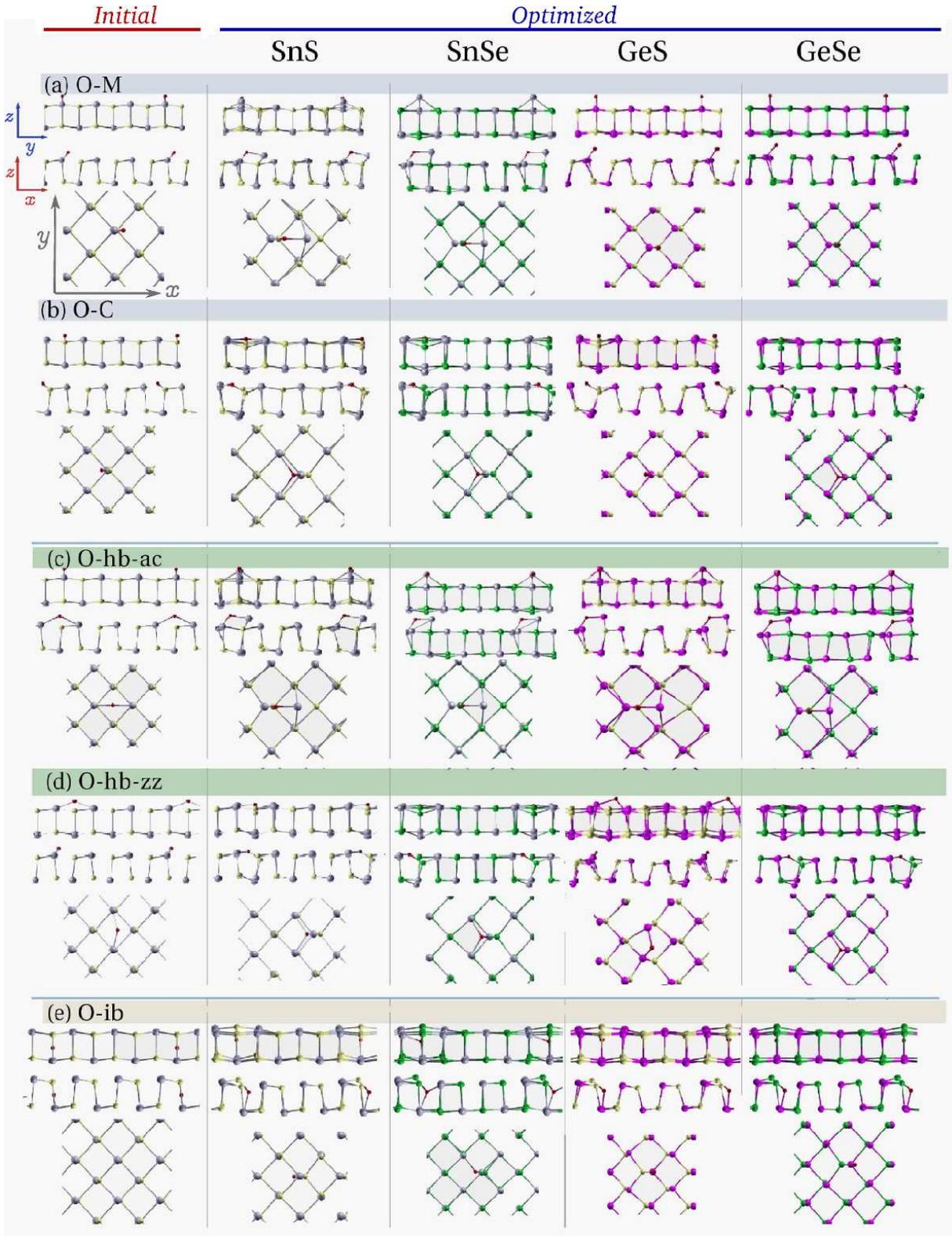}}
   \caption{\small (Color online) 
Structures of chemisorbed oxygen defects, where $M$=(Sn, Ge) and $C$=(S, Se). The structures of the initial positions are shown in the first column, with the initial positions of the oxygen, represented by the red atoms. The coordinate axes used in all systems are shown in the first row. The five different O positions are labeled: dangling oxygen defects (O-$M$ and O-$C$),  horizontal bridge defects,  one with O bonds parallel to the armchair $x$ direction (O-hb-ac), and a second one with O bonds along the zigzag $y$ direction (O-hb-zz), and interstitial oxygen (O-ib) is also discussed. The optimized layers for SnS, SnSe, GeS and GeSe are depicted in columns 2 to 5.}
   \label{O-MC}
 \end{figure*}

Group-IV monochalcogenides are isoelectronic with phosphorus, and their monolayer form assumes a corrugated structure very similar to phosphorene, with all atoms three-fold coordinated. The presence of two atomic species lowers the symmetry, and thus the bulk structure belongs to space group $Pnma-D^{16}_{2h}$, while black phosphorus is $Pnma-D^{18}_{2h}$. 
In the monolayer form, they also lose inversion symmetry in the perpendicular direction of the layers, which places them in the $Pn2_{1}m-C_{2\nu}^{7}$ space group.

Oxygen atoms can be adsorbed at different positions. 
As an initial step, we consider five configurations for isolated oxygen defects, 
derived from the models for oxygen defects in phosphorene considered in Ref.~\cite{PhysRevLett.114.046801}. 
The models can be divided into dangling, horizontal bridge and interstitial bridge configurations, and are shown in the first column in Fig.~\ref{O-MC}.

In dangling oxygen configurations the oxygen atom is bonded to only one lattice atom, borrowing two electrons from one of the lone pairs of $C$ or $M$.
Thus, in group-IV monochalcogenides there are two distinct dangling bond configurations,
unlike in phosphorene where there is only one.
These models are labeled O-$M$($C$)-$MC$, where $M$($C$) = Sn, Ge (S, Se) indicates the species oxygen is bonded to.
The two dangling oxygen structures are shown in Figs.~\ref{O-MC}~(a) and (b). 

We also consider two horizontal bridge configurations consisting of one oxygen positioned mid-way between two Sn(Ge) atoms of either two neighbor zigzag chains
(i.e. along the armchair direction), 
or the same zigzag chain, along the zigzag direction. These models are labeled O-hb-ac-$MC$ (Fig.~\ref{O-MC}(c)) and O-hb-zz-$MC$ 
(Fig.~\ref{O-MC}(d)), respectively.
In the bridge-type configurations the oxygen forms two single bonds to its nearest neighbors.
In the interstitial configuration, named O-ib-$MC$, the oxygen is initially placed at a bond center ie. between a $M$ and $C$ atoms, as shown in Fig.~\ref{O-MC}(e).

For all defects, the five initial structures (first column in Fig.~\ref{O-MC}) give rise to the respective optimized structures presented in columns 2 to 5 for SnS, SnSe, GeS and GeSe. 
The four materials show some variation in the final structures. 
After optimization, most defects result in significant distortion of the $MC$ structure in their neighborhood. Exceptions are dangling oxygen defects bonded to Ge in GeS and GeSe, 
for which the lattice remains little changed. 

In addition, not all structures are stable for all four materials.
Take as an example the SnS oxygen defects. 
In the horizontal bridge configuration O-hb-ac-SnS, oxygen atoms are initially bonded to two Sn atoms at equal distances O$-$Sn~$\simeq$~2.3~\AA{}. 
After optimization, oxygen pulls one of the Sn atoms to $h^{top}_{out}$=~0.80~\AA{} above the monolayer plane and along $x$ direction, 
towards to the second atom bonded to O. 
This is exactly the same structure adopted by the dangling oxygen O-Sn-SnS after optimization, 
as can be seen in Fig.~\ref{O-MC}. 
The dangling oxygen and bridge oxygen configurations, O-S-SnS and O-hb-zz-SnS, 
assume also very similar structures after optimization, with the lattices less affected by the introduction of the oxygen atoms. In this case, the sulfur in the O-S bond is pulled into the layer by about $h_{in}$=~0.37~\AA{} while the tin atom directly below it is pushed $h^{\rm bot}_{\rm out}$=~0.30~\AA{} out of the layer surface. For SnSe, the initial O-$M$ and O-hb-ac models also adopt the same final structures, as well O-$C$ and O-hb-zz. The main effect of the interstitial oxygen in the O-ib model is to push the chalcogen atom bonded to it slightly out of the monolayer plane. 

\begin{table}[t]
\centering
\renewcommand{\arraystretch}{1.4}
\begin{tabular}{m{1.8cm} m{1.2cm} m{1.2cm} m{1.2cm} m{1.4cm}}
\hline \hline
      & $h^{top}_{out}$ & $h_{in}$  & $h^{bot}_{out}$ & $h^{top}_{out}-ib$ \\ \hline
SnS   &    0.80         &  0.37     &    0.30         &      0.70          \\  
SnSe  &    1.52         &  0.68     &    0.57         &      0.61          \\ 
GeS   &    1.33         &  0.40     &    0.37         &      0.56          \\
GeSe  &    1.60         &  0.74     &    0.62         &      0.51          \\
\hline \hline
\end{tabular}
\caption{\small Structural parameters (in \AA{}) for the optimized structures.}
\label{h-opt-struc}
\end{table}

Along with an analysis of the structural changes, we investigate the energetic stability of the oxidized materials. The binding energy $E_b$ per oxygen atom is defined as:
 
\begin{equation}
E_b = E_{l+\rm O} - \left(E_{l} + N_{\rm O}\mu_{\rm O} \right)
\label{eb}
\end{equation}
where $ E_{l+\rm O}$ is the total energy of the defective layers, $E_{l}$ is the energy of the pristine layers (without the oxygen atom), $\mu_{\rm O}$ is the chemical potential of oxygen and $N_{\rm O}$ is the number of oxygen atoms per unit cell, which we have chosen as $N_{\rm O}$=1 in this work. A natural choice for $\mu_{\rm O}$, is the chemical potential of $\rm O_2$ molecules as the oxygen source, from which $\mu_{\rm O}$ is obtained by $E_{\rm O_2}$/2, where $E_{\rm O_2}$ is the total energy of an $\rm O_2$ molecule. Defined as in Eq.~\ref{eb}, a negative $E_b$ indicates that the defect formation is energetically favorable (exothermic reaction). The calculated $E_{b}$ for all materials and defect models are presented in Table~\ref{eb-tab}.

Both Sn chalcogenides have low oxygen binding energies, of about $-$0.7 eV, 
while Ge chalcogenides have oxygen binding energies around $-$1.2-1.4 eV.
This indicates that the latter are the most susceptible to oxidation.
Still, even in this case the absorption energy is lower than in phosphorene, where the binding energy for the dangling configuration is found to be $-2.08$~eV using a similar method. 

Overall, comparing the energies of the different defects across the four materials (Table~\ref{eb-tab}), we find that the lowest energy configuration is always the horizontal-bridge oxygen along the armchair direction, O-hb-ac.

\begin{table}[ht]
\centering
\renewcommand{\arraystretch}{1.4}
\begin{tabular}{m{2.4cm} m{1.4cm} m{1.4cm} m{1.4cm} m{1.4cm}}
\hline \hline
            &  SnS          &  SnSe         &    GeS   &     GeSe  \\ \hline
 O-$M$      &  -0.75        &   -0.75       &  -0.83   &  -0.69   \\
 O-$C$      &  -0.74        &   -0.51       &  -0.64   &  -0.42   \\
 O-hb-ac    &  -0.74$^{*}$  &   -0.75$^{*}$ &  -1.47   &  -1.22   \\ 
 O-hb-zz    &  -0.74$^{**}$ &   -0.51$^{**}$&  -0.77   &  -0.42   \\ 
 O-ic       &  -0.62        &   -0.56       &  -0.63   &  -0.46   \\
\hline \hline
\end{tabular}
\caption{\small Binding energies $E_{b}$ (eV) for chemisorbed oxygen atoms in monolayer monochalcogenides. For SnS and SnSe: ($^{*}$) same as O-$M$, ($^{**}$) same as O-$C$.}
\label{eb-tab}
\end{table}

\subsubsection{Electronic Properties}

\begin{figure*}[!htb]
    \centerline{
      \includegraphics[scale=0.90]{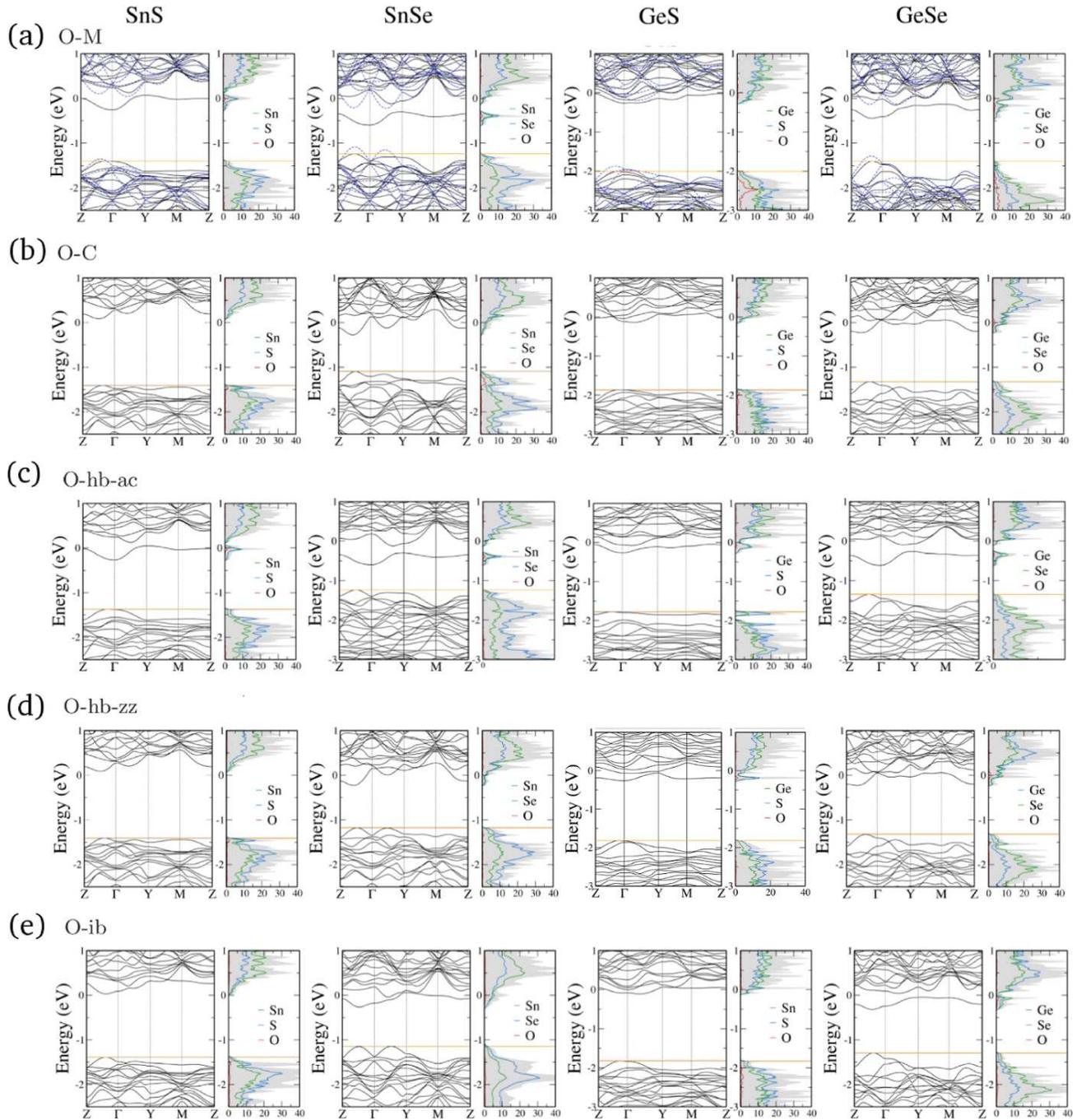}}
   \caption{(Color online) Electronic bands, total and partial DOS for (a) dangling O-$M$ oxygen, (b) dangling O-$C$ oxygen, (c) horizontal bridge O-hb-ac oxygen, (d) horizontal bridge O-hb-zz oxygen and (e) interstitial O-ib oxygen defects. For comparison, the electronic bands of a 3$\times$3 pristine unit cell is overlayed to the bandstructure of the O-$M$ models (first row) in dark blue dashed lines. The top of the valence band is marked by a horizontal orange line. The bands are shown in the same order as the corresponding relaxed structures in Fig.\ref{O-MC}.}
   \label{bands-O}
 \end{figure*}

Fig.~\ref{bands-O} shows the electronic bandstructures for all oxygen defect configurations. 
The total density of states (DOS) and the contribution of each atomic specie to the electronic states, i.e., the projected density of states (PDOS), are also presented. 
The electronic properties of monolayer monochalcogenides are appreciably affected by the introduction of oxygen. 
The characteristic valleys observed for the pristine structures~\cite{PhysRevB.92.085406} are strongly modified when they are exposed to oxygen. 
For all materials, the oxygen states hybridize with those of the $MC$ atoms. 

Let us first concentrate on the low energy horizontal bridge O-hb-ac-$MC$ defect configuration (Fig.~\ref{bands-O}(c)).
For SnS, SnSe and GeSe, empty level states are introduced in the gap at 75~meV, 178~meV and 148~meV, respectively, from the conduction band minima, localized along the $\Gamma$-Y lines of the Brillouin zone (BZ). 
In GeS, an occupied state is formed 134~meV above the top of the valence band, while an empty state is calculated at $\approx$~25~meV below the lowest conduction band. 
An analysis of the PDOS shows that the acceptor and donor bands are mainly formed by the states of the $MC$ atoms, and only a minor contribution from the oxygen atom (less than $\sim$~8\% in SnS). The horizontal bridge defects O-hb-zz-$MC$, in contrast, do not introduce gap states and are, therefore, electrically neutral for all materials.  

\begin{figure}[!htb]
    \centerline{
  \includegraphics[scale=0.22]{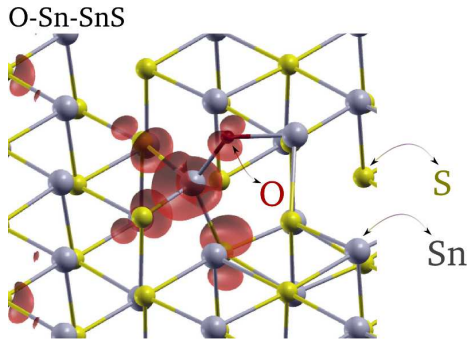}}
   \caption{\small (Color online) Charge density contribution to the defective state of the O-Sn-SnS model. The charge density distributions for the other compounds (SnSe, GeS and GeSe) are very similar and are not shown for simplicity.}
   \label{1-O-Sn-SnS.rho}
 \end{figure}

The electronic bands, total and partial density of states for the monolayers with dangling oxygen configurations are presented in Figs.~\ref{bands-O}~(a) and (b). When the oxygen is bonded to a chalcogen atom (O-$C$ structures), no gap states are introduced. 
On the other hand, when oxygen is bonded to a group-IV atom (O-$M$ structures), the conduction band is perturbed and acceptor levels are introduced. 
The gap state is localized mainly on the Sn or Ge atoms directly bonded to the oxygen. Fig.~\ref{1-O-Sn-SnS.rho} shows the charge density contribution of such defective in-gap state in SnS. 

For the interstitial bridge defects, O-ib-$MC$, the top of the valence band is less affected, if compared to the other defect models, while the conduction band is still strongly modified in comparison to the pristine layers. In particular, for GeSe an acceptor state is formed a few meV bellow the conduction band minimum.

 
Activation energies can be estimated by the Marker Method (MM), as detailed in Ref.~\cite{marker}. This method allows us to estimate energy levels comparing ionization energies of defective systems in different charge states, referent to a known {\it marker} system. A natural choice as reference system is the pristine (undefective) material, when other appropriate markers, for which reliable experimental data exists, are not available. With these considerations, its shown~\cite{marker} that the electrical levels of an unknown system can be defined with respect to a known marker system by means of their total energies.  
We need to emphasize that the MM is most reliable for comparison between chemically (and structurally) similar systems.
In the case of 2D materials modeled with periodic boundary conditions,
the marker method is an efficient way to cancel, in a good approximation, 
the energy resulting from the spurious electrostatic interaction between neighbouring cells\cite{PhysRevB.89.081406}.

For the models discussed up to now, we focus on calculate activation energies for the O-$M$ and low energy O-hb-ac systems, which present in-gap defective acceptor bands for all the monochalcogenides investigated here. The activation energies of the defective states are calculated from the differences in electron affinities:
\begin{equation}
I_D - I_m = [E_D(0) - E_D(q)] - [E_m(0) - E_m(q)]
\label{ae}
\end{equation} 
where $E(q)$ is the energy of the supercell in charge state $q=\lbrace-,+\rbrace$, and the sub-indices $m$($D$) refer to pristine (defective) systems. Charged systems with an extra electron (q = $-$1) or a missing electron (q = +1) are considered for the pristine and defective monolayers with acceptor or donor states, respectively. The total energies used in Eq.~\ref{ae} are computed from GGA-PBE exchange-correlation functional. The calculated results are summarized in Table~\ref{ae-tab}.

All acceptor levels are deep, lying in the upper half of the bandgap. The O-hb-ac defect in GeS presents the shallower state, with activation energy of 50~meV. GeS is also the only material which presents a donor defective state. Such state is deep, with an activation energy of 100~meV.

\begin{table}[ht]
\centering
\renewcommand{\arraystretch}{1.4}
\begin{tabular}{m{2cm} m{1.2cm} m{1.2cm} m{1.2cm} m{1.2cm} m{0.8cm}}
 \hline \hline 
            &  SnS    &  SnSe  & GeS    &  GeSe &        \\ \hline 
O-$M$       &  0.17   &  0.31  &  0.11  &  0.22 & ($-$/0)  \\
O-hb-ac     &  0.17   &  0.31  &  0.05  &  0.33 & ($-$/0)  \\
            &  -      &   -    &  0.10  &   -   & (0/+)  \\
V$_{C}$     &  0.21   &   -    &  0.07  &   -   & (0/+)  \\
 \hline \hline
\end{tabular}
\caption{\small Activation energies (in eV) of the defects in O-$M$ and O-hb-ac adsorbed oxygen systems and of the chalcogen vacancies V$_{C}$. For the adsorbed O models, the acceptor levels ($-$/0) are deep and located in the upper half of the band gap. The V$_{C}$ vacancies present donor states (0/+) only for suphides, being shallower in GeS than in SnS.}
\label{ae-tab}
\end{table}

\subsection{Vacancies and Substitutional Oxygen}
\label{V-O-defects}

Besides the oxygen defects discussed in the previous section, intrinsic defects is another class of dominant defects present in 2D materials. 
Bulk SnS, for instance, is characterized by an intrinsic p-type conductivity due to typical acceptor states formed by Sn vacancies ($\rm V_{Sn}$)~\cite{apl.100.032104}. S vacancies ($\rm V_{S}$) can also be formed under appropriate Sn-rich conditions as well substitutional oxygen at S sites ($\rm O_{S}$). Experimental studies also indicate the presence of vacancies in single-crystal GeSe nanosheets, and discuss the role of the resulting defective states in the photoresponse of this material~\cite{am402550s}. In this section, we investigate intrinsic defects in the monolayers of group-IV monochalcogenides. We discuss the energetic of the systems with introduction of four different types of defects: two vacancies, V$_{M}$ and V$_{C}$, and two substitutional oxygen defects, O$_{M}$ and O$_{C}$.

\begin{figure}[!htb]
    \centerline{
  \includegraphics[scale=0.45]{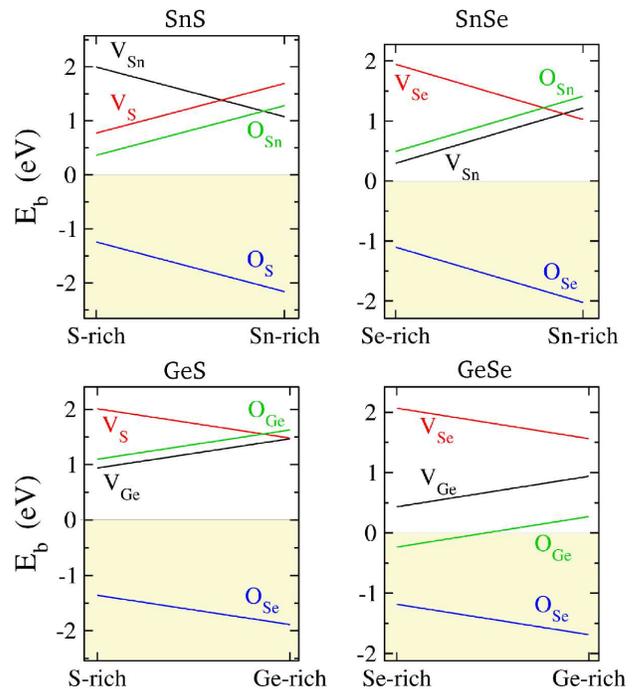}}
   \caption{\small (Color online) Formation energies $E_{f}$ (eV) for vacancies and substitutional oxygen defects for $M$-rich ($M$ = Sn, Ge) and $C$-rich ($C$ = S, Se) conditions. All the vacancies present positive formation energies, an indication that their formation process is endothermic. On the other side, oxygen substitutional at the chalcogen sites ($\rm O_{S}$ and $\rm O_{Se}$ defects), are energetically favorable to occur in these systems, given their remarkable negative $E_{b}$ values.}
   \label{vac-subst}
 \end{figure}

Formation energies ($E_f$) are calculated using
\begin{equation}
E_f = E_{\rm def} - \left(N_{M}\mu_{M} + N_{C}\mu_{C} + N_{\rm O}\mu_{\rm O} \right)
\label{eb2}
\end{equation}
where $E_{def}$ is the total energy of the defective structure, $\mu_{i}$ and $N_{i}$ are the chemical potential and number of atoms of $i$ type. The chemical potentials for $M$ and $C$ ($\mu_{M}$ and $\mu_{C}$, respectively)  for Ge and Sn-rich conditions are taken from the diamond structure of these elements. The chemical potentials for Se and S-rich environments are calculated using the molecular crystal (R$\overline{3}$ phase) Se$_6$ and the S$_8$ molecule~\cite{chemsci.7.1082}. The formation energy ($E_{f}$) interval is presented in Fig.~\ref{vac-subst}, where, as in the previous section, our definition of formation energy yields  negative values for exoenergetic processes.

\begin{figure*}[!htb]
    \centerline{
  \includegraphics[scale=0.65]{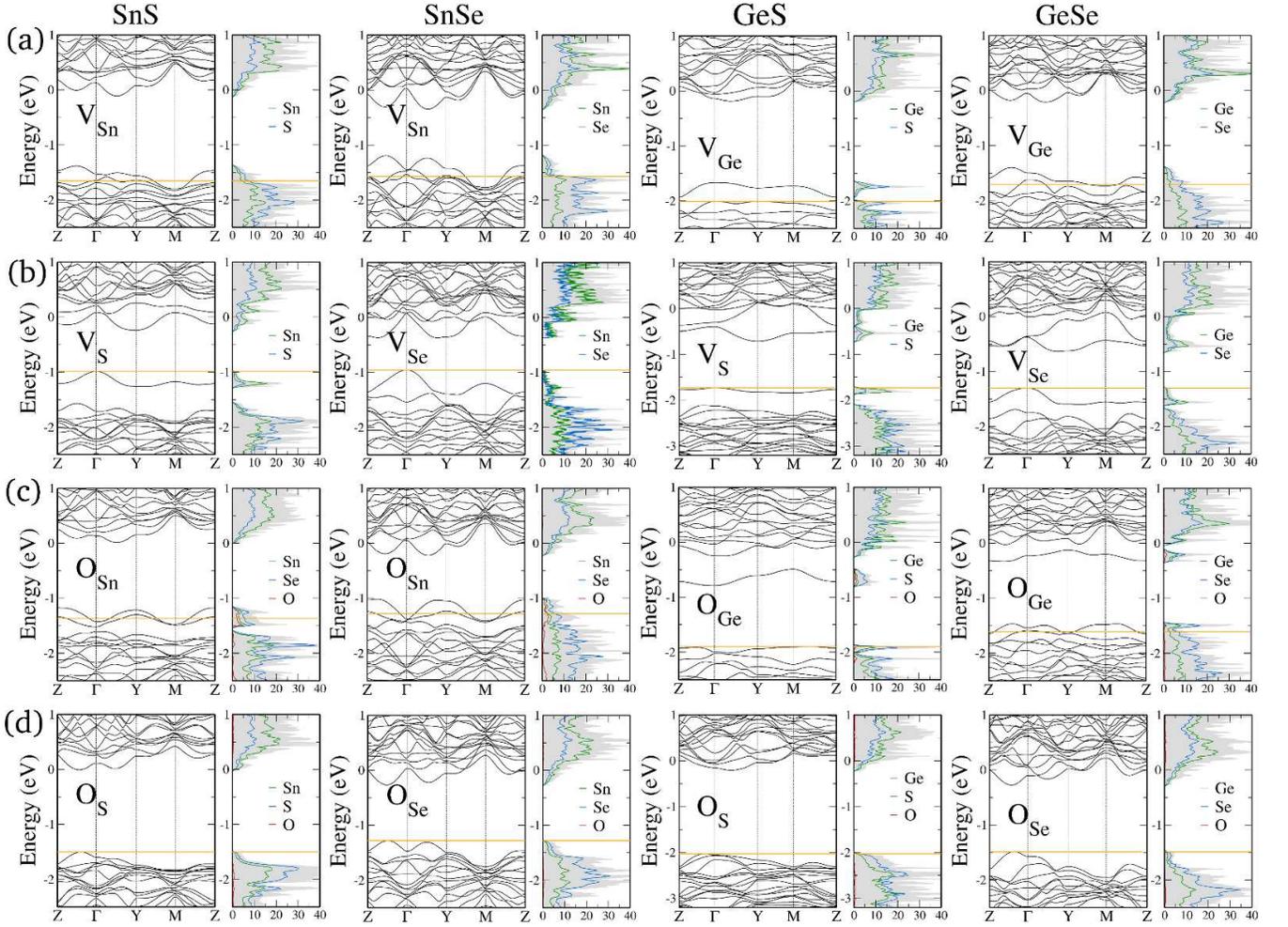}}
   \caption{\small (Color online) Electronic bands, DOS and PDOS for monolayers with (a) vacancies of group-IV elements (Sn, Ge), (b) vacancies of chalcogens  (S, Se), (c) substitutional O defects in group-IV elements (Sn, Ge), (d) substitutional O defects in chalcogen atoms (S, Se). The top of the valence band is marked by a horizontal orange line. The bands are shown in the same order as the corresponding relaxed structures in Fig.\ref{O-MC}.}
   \label{bands-V-O}
 \end{figure*}

The first marked result is the indication of stability of all O-substitutional defects at the chalcogen sites, given by the negative values of the $\rm O_{S}$ and $\rm O_{Se}$ defects. In addition to the $\rm O_{S/Se}$, the only defect model that presents negative formation energy is the $\rm O_{Ge}$ in GeSe, under Se-rich condition. All vacancies, as well the remainder substitutional oxygen $\rm O_{Sn/Ge}$, are endothermic processes and, at least in the growth environments considered here, should not be favorable to occur in the single-layer of these materials. 

Formation energies of vacancies and oxygen substituting for Sn/Ge are positive, however the formation energies of oxygen replacing S/Se are negative. Thus, the reaction $\rm V_{S/Se}+\frac{1}{2} O_{2}\rightarrow O_{S/Se}$ is energetically favorable.

\subsubsection{Electronic properties}

Inspection of the electronic structure of the different types of vacancy defects indicates
that Sn/Ge vacancies remove valence states, acting as shallow acceptors and displacing the Fermi level below the valence band top, as shown in Fig.~\ref{bands-V-O}(a). In the case of chalcogen vacancies, perturbed valence and conduction bands appear in the selenides. A similar perturbed state has been predicted for bulk GeSe with the same type of defect~\citep{am402550s}. For the sulphides, S vacancies introduce deep donor states, localized at Sn and Ge atoms in the vicinity of the missing S atom (Fig.~\ref{bands-V-O}b). Activation energies of the $\rm V_{S}$ states are presented in Table~\ref{ae-tab} and show that the defect state is shallower in GeS (70~meV) than in SnS (210~meV).

\begin{figure}[!htb]
    \centerline{
  \includegraphics[scale=0.22]{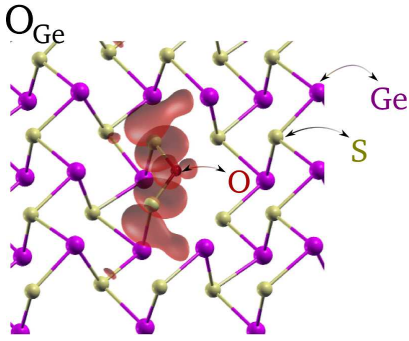}}
   \caption{\small (Color online) Isosurface of charge density for the defect band introduced in the $\rm O_{Ge}$ substitutional defect in GeS. The plot shows the localized nature of the bands in the region around the oxygen and first S and Ge neighbouring atoms.}
   \label{o-ge-ges-rho}
 \end{figure}

$\rm O_{Sn/Ge}$ is also shallow acceptor in all cases except GeS, where it introduces a deep state instead (Fig.\ref{bands-V-O}(c)). This deep state is highly localized on the oxygen atom and on the S atom bonded to it, as shown by the charge density distribution of the defect state, plotted in Fig.~\ref{o-ge-ges-rho}.

In contrast, substitutional $\rm O_{S/Se}$ defects are isoelectronic with the pristine structure and do not introduce gap states for most of the compounds, as shown in Fig.~\ref{bands-V-O}d, leaving the bandstructure mostly unaffected.

Given the lack of experimental results on the properties of monolayer monochalcogenides, we establish a comparison with bulk and few-layer material.
In Ref.~\citep{am402550s}, for instance, photoresponse analysis of single-crystal GeSe nanosheets shows a slow decay time, which is attributed to defect states created in the samples upon chemical process or light and heat application in the fabrication process. 
Indeed, a previous first-principles study of defects in bulk GeSe, shows the presence of middle gap states for vacancies or interstitial atoms. 
The energetic preference of oxygen to be incorporated into the layers, occupying the chalcogen sites (S and Se atoms), at the two limits of chemical potentials
has also been reported by a previous study of surface passivation of bulk SnS~\cite{jap.115.173702}. 

\section{Conclusions}
\label{conclusions}

Point defects in group-IV monochalcogenides monolayers - SnS, SnSe, GeS and GeSe - are investigated using first principles density functional theory calculations. 
Energetic and structural analysis of five different models for chemisorbed oxygen atoms, reveals a better resistance of these materials to oxidation if compared to their isostructural partner, phosphorene. Amongst all monochalcogenides, GeS is the most prone to oxidation, as it presents larger binding energies for four of the five models investigated. 

Electronic structure calculations show that the most stable oxygen configurations have deep acceptor states, 
and so do the chalcogen vacancies. 
However, oxygen substitution leads to neutral defects which preserve the electronic structure of the pristine material.
Substitutional oxygen forms spontaneously at the chalcogen sites in the presence of chalcogen vacancies and oxygen. 
This indicate that annealing/laser healing of vacancy defects will be effective in removing gap states in group-IV monochalcogenides, 
as was found for TMDCs.\cite{junpeng}

In contrast, Sn/Ge vacancies are shallow acceptors,
and therefore are expected to confer $p$-type character to chalcogen-rich material.
In this case, annealing in oxygen is not expected to be an effective passivation technique.

\section*{Acknowledgements}
This work was supported by the National Research Foundation, Prime Minister Office, Singapore,
under its Medium Sized Centre Programme and CRP
award ``Novel 2D materials with tailored properties: beyond graphene" (Grant number R-144-000-295-281).
The first-principles calculations were carried out on the GRC high-performance computing facilities.

\label{Bibliography}
%
%
 \bibliographystyle{unsrtnat} 
 \bibliography{Bibliography} 

\end{document}